\documentclass{elsart}

\usepackage{amssymb,amsmath}
\newtheorem{theorem}{Theorem}[section]
\newtheorem{definition}{Definition}[section]
\newtheorem{remark}{Remark}[section]
\newtheorem{lemma}{Lemma}[section]

\begin{document}

\begin{frontmatter}



\title{Quadratic metric-affine gravity}


\author{Dmitri Vassiliev}
\ead{D.Vassiliev@bath.ac.uk}
\ead[url]{http://www.bath.ac.uk/\~{}masdv/}

\address
{Department of Mathematical Sciences, University of Bath, Bath BA2 7AY, UK}

\begin{abstract}
We consider spacetime to be a connected real 4-manifold equipped with a
Lorentzian metric and an affine connection. The 10 independent
components of the (symmetric) metric tensor and the 64 connection
coefficients are the unknowns of our theory.
We introduce an action which is
quadratic in curvature and study the resulting system of
Euler--Lagrange equations.
In the first part of the paper we look for Riemannian solutions,
i.e. solutions whose connection is Levi-Civita.
We find two classes of Riemannian solutions: 1)~Einstein spaces, and
2)~spacetimes with metric of a pp-wave and parallel Ricci curvature.
We prove that
for a generic quadratic action these are the only Riemannian solutions.
In the second part of the paper we look for non-Riemannian solutions.
We define the notion of a ``Weyl pseudo\-instanton''
(metric compatible spacetime whose curvature is purely Weyl)
and prove that a Weyl pseudo\-instanton is a solution of our field equations.
Using the pseudo\-instanton
approach we construct explicitly a non-Riemannian solution
which is a wave of torsion in Minkowski space.
We discuss the possibility of using this non-Riemannian
solution as a mathematical model for the graviton or the neutrino.
\end{abstract}

\end{frontmatter}

\newpage
\tableofcontents
\newpage

\section{Mathematical model}
\label{model}

We consider spacetime to be a connected real
4-manifold $M$ equipped with a Lorentzian metric $g$
and an affine connection $\Gamma$.
The 10 independent
components of the (symmetric) metric tensor $g_{\mu\nu}$
and the 64 connection coefficients ${\Gamma^\lambda}_{\mu\nu}$
are the unknowns of our theory.
This approach is known as metric-affine gravity.
Its origins lie in the works of authors such as
\'E.~Cartan,
A.S.~Eddington,
A.~Einstein,
T.~Levi-Civita,
E.~Schr\"odinger
and H.~Weyl.
A review of the more recent
work in this area can be found in \cite{hehlreview}.

We define our action as
\begin{equation}
\label{action}
S:=\int q(\!R)
\end{equation}
where $q$ is an $\mathrm{O}(1,3)$-invariant
quadratic form on curvature $R\,$.
Independent variation of the metric
$g$ and the connection $\Gamma$ produces Euler--Lagrange
equations which we will write symbolically as
\begin{eqnarray}
\label{eulerlagrangemetric}
\partial S/\partial g&=&0,
\\
\label{eulerlagrangeconnection}
\partial S/\partial\Gamma&=&0.
\end{eqnarray}
Our objective is the study of the
combined system of field equations
(\ref{eulerlagrangemetric}), (\ref{eulerlagrangeconnection}).
This is a system of $10+64$
real nonlinear partial differential equations
with $10+64$ real unknowns.

Our motivation comes from Yang--Mills theory. The Yang--Mills action
for the affine connection is a special case of (\ref{action}) with
\begin{equation}
\label{YMq}
q(\!R)=q_{\text{YM}}(\!R):=
R^\kappa{}_{\lambda\mu\nu}\,
R^\lambda{}_\kappa{}^{\mu\nu}\,.
\end{equation}
With this choice of $q$ equation (\ref{eulerlagrangeconnection}) is
the Yang--Mills equation for the affine connection.
There is a substantial bibliography devoted to the study of the system
(\ref{eulerlagrangemetric}), (\ref{eulerlagrangeconnection})
in the special case (\ref{YMq}); see, for example, references in
\cite{pseudo}. Without going into the details of the historical development
of the subject let us mention the contributions of
C.N.~Yang \cite{yang} and E.W.~Mielke \cite{mielkepseudoparticle}
who showed, respectively, that Einstein spaces satisfy equations
(\ref{eulerlagrangeconnection}) and (\ref{eulerlagrangemetric}).

The idea of using a quadratic action in General Relativity goes
back to H.~Weyl, see end of his paper \cite{weylquadraticaction}.
Weyl also pointed out that such an action should contain
all possible invariant quadratic combinations of curvature,
say, the square of Ricci curvature, the square of scalar curvature, etc.
It turns out (see Appendix \ref{appendixa}) that in the metric-affine
setting curvature has 11 irreducible pieces.
There are (see Appendix \ref{appendixb}) 16 ways of squaring these
irreducible pieces to a scalar.
The reason why the number of different quadratic combinations
is greater than the number of irreducible pieces is that
some of the irreducible pieces are isomorphic.
The general formula for an $\mathrm{O}(1,3)$-invariant
quadratic form on curvature is given in
Lemma \ref{lemma b1}.

\begin{definition}
\label{definition of riemannian}
We call a spacetime $\{M,g,\Gamma\}$
\emph{Riemannian}
if the connection is Levi-Civita
(i.e. ${\Gamma^\lambda}_{\mu\nu}=
\genfrac{\{}{\}}{0pt}{}{\lambda}{\mu\nu}$),
and \emph{non-Riemannian}
otherwise.
\end{definition}

\begin{remark}
\label{remark on riemannian}
The word ``Riemannian'' has a different meaning in mathematics
and theoretical physics.
In mathematical literature
the connection is usually Levi-Civita by default
and ``Riemannian'' indicates that the metric is definite,
whereas in theoretical physics literature
the metric is usually Lorentzian by default (as it is in our paper)
and ``Riemannian'' indicates that the connection is Levi-Civita.
In Definition \ref{definition of riemannian}
we adopt the theoretical physics terminology.
\end{remark}

\begin{remark}
\label{remark on euclidean}
We call definite metrics ``Euclidean'', and
non-degenerate metrics of arbitrary signature ``pseudo-Euclidean''.
\end{remark}

The aim of this paper is to study the field equations
(\ref{eulerlagrangemetric}), (\ref{eulerlagrangeconnection}), so as to find
\begin{itemize}
\item
\emph{all} Riemannian solutions, and
\item
\emph{some} non-Riemannian solutions.
\end{itemize}

The paper has the following structure.
In Section \ref{field} we write down explicitly the field
equations
(\ref{eulerlagrangemetric}), (\ref{eulerlagrangeconnection})
in the Riemannian case.
In Sections \ref{type1}--\ref{type3} we construct three types of
Riemannian solutions.
In Section \ref{uniqueness} we prove a uniqueness theorem
stating that for a generic quadratic action
solutions from Sections \ref{type1}--\ref{type3} are the only
Riemannian solutions;
this uniqueness theorem is the main result of our paper.
In subsequent sections we look for
non-Riemannian solutions, and succeed
(Section \ref{Non-Riemannian pseudoinstantons})
in constructing explicitly one particular non-Riemannian solution.
We discuss our results in Section \ref{Discussion}.
Finally,
Appendices \ref{appendixa}--\ref{appendixc} contain
statements and proofs of some
auxiliary mathematical facts.

\section{Notation}
\label{notation}

Our notation follows \cite{King and Vassiliev,pseudo}. In particular,
we denote local coordinates by $x^\mu$, $\mu=0,1,2,3$,
and write $\partial_\mu:=\partial/\partial x^\mu$.
We define the covariant derivative of a vector function as
$\nabla_{\mu}v^\lambda:=\partial_\mu v^\lambda
+{\Gamma^\lambda}_{\mu\nu}v^\nu$,
torsion as
${T^\lambda}_{\mu\nu}:=
{\Gamma^\lambda}_{\mu\nu}-{\Gamma^\lambda}_{\nu\mu}\,$,
curvature as
${R^\kappa}_{\lambda\mu\nu}:=
\partial_\mu{\Gamma^\kappa}_{\nu\lambda}
-\partial_\nu{\Gamma^\kappa}_{\mu\lambda}
+{\Gamma^\kappa}_{\mu\eta}{\Gamma^\eta}_{\nu\lambda}
-{\Gamma^\kappa}_{\nu\eta}{\Gamma^\eta}_{\mu\lambda}\,$,
Ricci curvature as
$Ric_{\lambda\nu}:={R^\kappa}_{\lambda\kappa\nu}\,$,
scalar curvature as $\mathcal{R}:=Ric^\lambda{}_\lambda\,$,
and trace-free Ricci curvature as
$\mathcal{R}ic_{\lambda\nu}:=
Ric_{\lambda\nu}-\frac14g_{\lambda\nu}\mathcal{R}\,$.
We denote Weyl curvature by $\mathcal{W}=R^{(10)}$
(see also Appendix \ref{appendixa}).
Given a scalar function $f:M\to\mathbb{R}$ we write for brevity
$\displaystyle\int f:=\int_Mf\,\sqrt{|\det g|}
\,\d x^0\d x^1\d x^2\d x^3$
where $\,\det g:=\det(g_{\mu\nu})\,$.
The totally antisymmetric quantity is
denoted by $\varepsilon_{\kappa\lambda\mu\nu}$.
The Christoffel symbol is
$\genfrac{\{}{\}}{0pt}{}{\lambda}{\mu\nu}:=
\frac12g^{\lambda\kappa}
(\partial_\mu g_{\nu\kappa}
+\partial_\nu g_{\mu\kappa}
-\partial_\kappa g_{\mu\nu})$.


\section{Field equations in the Riemannian case}
\label{field}

When looking for Riemannian solutions we need to \emph{specialise}
our field equations
(\ref{eulerlagrangemetric}), (\ref{eulerlagrangeconnection})
to the Levi-Civita connection.
We will write the resulting equations symbolically as
\begin{eqnarray}
\label{eulerlagrangemetricLC}
\left.\partial S/\partial g\,\right|_{\text{L-C}}&=&0,
\\
\label{eulerlagrangeconnectionLC}
\left.\partial S/\partial\Gamma\,\right|_{\text{L-C}}&=&0.
\end{eqnarray}
It is important to understand the logical sequence involved in
the derivation of equations
(\ref{eulerlagrangemetricLC}), (\ref{eulerlagrangeconnectionLC}):
we set
${\Gamma^\lambda}_{\mu\nu}=
\genfrac{\{}{\}}{0pt}{}{\lambda}{\mu\nu}\,$
\emph{after} the variations
of the metric and the connection have been carried out.

Equations
(\ref{eulerlagrangemetricLC}), (\ref{eulerlagrangeconnectionLC})
are equations for the unknown metric
in the usual, Riemannian, setting.
In the Riemannian case curvature has only 3 irreducible
pieces, so the LHS's of
(\ref{eulerlagrangemetricLC}), (\ref{eulerlagrangeconnectionLC})
can be expressed via
scalar curvature $\mathcal{R}$,
trace-free Ricci curvature $\mathcal{R}ic$,
and Weyl curvature $\mathcal{W}$.
Lengthy but straightforward calculations give the
following explicit representation for equations
(\ref{eulerlagrangemetricLC}), (\ref{eulerlagrangeconnectionLC}):
\begin{eqnarray}
\label{eulerlagrangemetricLCexplicit}
\!\!\!\!\!\!\!\!
d_1\mathcal{W}^{\kappa\lambda\mu\nu}\mathcal{R}ic_{\kappa\mu}
+d_2\,\mathcal{R}\,\mathcal{R}ic^{\lambda\nu}
+d_3\left(
\mathcal{R}ic^{\lambda\kappa}\mathcal{R}ic_\kappa{}^\nu
-\frac14g^{\lambda\nu}
\mathcal{R}ic_{\kappa\mu}\mathcal{R}ic^{\kappa\mu}\right)&=&0,
\\
\label{eulerlagrangeconnectionLCexplicit}
\!\!\!\!\!\!\!\!
d_4g_{\kappa\mu}\partial_\lambda\mathcal{R}
-d_5g_{\lambda\mu}\partial_\kappa\mathcal{R}
+d_6\nabla_{\!\lambda}\mathcal{R}ic_{\kappa\mu}
-d_7\nabla_{\!\kappa}\mathcal{R}ic_{\lambda\mu}
&=&0,
\end{eqnarray}
where
\begin{equation}
\label{formluae for ds}
\!
\begin{array}{lll}
d_1=b_{912}-b_{922}+b_{10},&
d_2=-b_1-\frac{b_{911}}4+\frac{b_{912}}6+\frac{b_{922}}{12},&
d_3=b_{922}-b_{911},\\
d_4=-b_1+\frac{b_{912}-b_{922}}4+\frac{b_{10}}{12},&
d_5=-b_1+\frac{b_{912}-b_{911}}4+\frac{b_{10}}{12},&
{}\\
d_6=b_{912}-b_{911}+b_{10},&
d_7=b_{912}-b_{922}+b_{10},&
{}
\end{array}
\end{equation}
the $b$'s being the coefficients from formula (\ref{appendixb3}).
Observe that the LHS of (\ref{eulerlagrangemetricLCexplicit}) is
trace-free. This is a consequence of the conformal invariance of
our action~(\ref{action}), see also Remark 1.1 in \cite{pseudo}.

The LHS's of equations
(\ref{eulerlagrangemetricLCexplicit})
and
(\ref{eulerlagrangeconnectionLCexplicit})
are the components of the tensors $A$ and $B$ from the formula
$\displaystyle\delta S=\int(2A^{\lambda\nu}\,\delta g_{\lambda\nu}
+2B^{\kappa\mu}{}_\lambda\,\delta\Gamma^\lambda{}_{\mu\kappa})$.
Here $\delta g$ and $\delta\Gamma$ are the (independent)
variations of the metric and the connection,
and $\delta S$ is the resulting variation of the action.
In (\ref{eulerlagrangeconnectionLCexplicit}) we lowered
the first two indices of $B$ to make the expression easier to read.
The same conventions were used in equations (25) and (26) of \cite{pseudo}.

In deriving explicit formulae for tensors $A$ and $B$
we simplified our calculations by adopting the following argument.
Formula (\ref{appendixb3}) can be rewritten as
\begin{multline*}
q(\!R)=b_1\mathcal{R}^2
+\sum_{l,m=1}^2 b_{9lm}(\mathcal{S}^{(l)},\mathcal{S}^{(m)})
+b_{10}(R^{(10)},R^{(10)})_{\mathrm{YM}}+\ldots
\\
=b_1\mathcal{R}^2
+\sum_{l,m=1}^2 b_{9lm}(\mathcal{R}ic^{(l)},\mathcal{R}ic^{(m)})
+b_{10}(R^{(10)},R^{(10)})_{\mathrm{YM}}+\ldots
\end{multline*}
where by $\ldots$ we denote terms which do not contribute to
$\delta S$ when we start our variation from a Riemannian spacetime.
Recall that the $\mathcal{R}ic^{(l)}$ are defined in accordance with
(\ref{appendixa5}).
Put
\[
Ric_\pm=\frac{Ric^{(1)}\pm Ric^{(2)}}2\,,
\qquad
\mathcal{R}ic_\pm=Ric_\pm-\frac14g\,\operatorname{tr}\!Ric_\pm
=\frac{\mathcal{R}ic^{(1)}\pm\mathcal{R}ic^{(2)}}2\,.
\]
Note that the tensor $Ric_+$ is trace-free,
and that in the Riemannian case we get $Ric_+=0$, $Ric_-=Ric$.
Our quadratic form can now be rewritten as
\begin{multline*}
q(\!R)
=b_1\mathcal{R}^2
+b_{10}(R^{(10)},R^{(10)})_{\mathrm{YM}}
+(b_{911}-2b_{912}+b_{922})(\mathcal{R}ic_-,\mathcal{R}ic_-)
\\
+2(b_{911}-b_{922})(Ric,Ric_+)+\ldots
\\
=\sum_{j=1}^3c_j(R^{(j)},R^{(j)})_{\mathrm{YM}}
+2(b_{911}-b_{922})(Ric,Ric_+)+\ldots
\end{multline*}
where
\begin{equation}
\label{formluae for cs}
c_1=-\frac12(b_{911}-2b_{912}+b_{922}),\qquad
c_2=-6b_1,\qquad
c_3=b_{10},
\end{equation}
and the $R^{(j)}$s are the irreducible pieces of curvature
labelled in accordance with \cite{pseudo};
note that the labelling of irreducible pieces in \cite{pseudo}
differs from that in the current paper.
The variation of
$\displaystyle\int\sum_{j=1}^3c_j(R^{(j)},R^{(j)})_{\mathrm{YM}}$
is given by the LHS's of formulae (25), (26) of \cite{pseudo}.
Thus, the problem reduces to computing the variation
of $\displaystyle\int(Ric,Ric_+)$. The latter turns out to be
\begin{multline*}
\delta\int(Ric,Ric_+)=
\\
\int
\left(
\frac12
\,\mathcal{W}^{\kappa\lambda\mu\nu}\mathcal{R}ic_{\kappa\mu}
-\frac16
\,\mathcal{R}\,\mathcal{R}ic^{\lambda\nu}
-\left(
\mathcal{R}ic^{\lambda\kappa}\mathcal{R}ic_\kappa{}^\nu
-\frac14g^{\lambda\nu}
\mathcal{R}ic_{\kappa\mu}\mathcal{R}ic^{\kappa\mu}\right)
\right)\delta g_{\lambda\nu}
\\
+\int
\left(
\frac18(g_{\kappa\mu}\partial_\lambda\mathcal{R}
+g_{\lambda\mu}\partial_\kappa\mathcal{R})
-\frac12(\nabla_\lambda\mathcal{R}ic_{\kappa\mu}
+\nabla_\kappa\mathcal{R}ic_{\lambda\mu})
\right)\delta\Gamma^{\lambda\mu\kappa}\,.
\end{multline*}
Consequently, the constants $d_1,\ldots,d_7$ appearing in
(\ref{eulerlagrangemetricLCexplicit}),
(\ref{eulerlagrangeconnectionLCexplicit})
are expressed via the constants $c_1,c_2,c_3$ and
$b_{911}-b_{922}$ as
\begin{equation}
\label{formluae for ds via cs}
\!\!\!\!\!\!\!\!\!\!
\begin{array}{lll}
d_1=c_1+c_3+\frac{b_{911}-b_{922}}2,&
d_2=\frac{c_1+c_2-b_{911}+b_{922}}6,&
d_3=b_{922}-b_{911},\\
d_4=\frac{c_1}4+\frac{c_2}6+\frac{c_3}{12}+\frac{b_{911}-b_{922}}8,&
d_5=\frac{c_1}4+\frac{c_2}6+\frac{c_3}{12}-\frac{b_{911}-b_{922}}8,&
{}\\
d_6=c_1+c_3-\frac{b_{911}-b_{922}}2,&
d_7=c_1+c_3+\frac{b_{911}-b_{922}}2.&
{}
\end{array}
\end{equation}
Substituting (\ref{formluae for cs})
into (\ref{formluae for ds via cs}) we arrive at
(\ref{formluae for ds}).

\section{Riemannian solutions of type 1}
\label{type1}

\begin{definition}
An \emph{Einstein space} is a Riemannian spacetime
with $Ric=\Lambda g$ where $\Lambda$ is some real
``cosmological'' constant.
\end{definition}

For an Einstein space
$\partial\mathcal{R}\equiv0$ and $\mathcal{R}ic\equiv0$, so
equations (\ref{eulerlagrangemetricLCexplicit}),
(\ref{eulerlagrangeconnectionLCexplicit}) are clearly satisfied.
We call Einstein spaces
\emph{Riemannian solutions of type 1.}

\section{Riemannian solutions of type 2}
\label{type2}

A metric of the form
\begin{equation}
\label{metric of a pp-wave}
g_{\mu\nu}\d x^\mu\d x^\nu=
\,2\,\d x^0\d x^3-(\d x^1)^2-(\d x^2)^2 +f(x^1,x^2,x^3)\,(\d x^3)^2
\end{equation}
is called a \emph{metric of a pp-wave,}
see Section 21.5 in \cite{exact solutions}.
Such metrics were introduced by Peres \cite{peres}
and have since been widely used
in General Relativity. The remarkable property of the metric
(\ref{metric of a pp-wave}) is that the corresponding
curvature tensor $R$ is linear in $f$.

There are differing views on what the ``pp'' stands for.
According to \cite{exact solutions} ``pp'' is an abbreviation for
``plane-fronted gravitational waves with parallel rays''. According
to Peres himself \cite{peresweb} ``pp'' is an abbreviation for
``plane polarized gravitational waves ''.

\begin{definition}
\label{definition 1 of a pp-space}
A \,\emph{pp-space} is a Riemannian spacetime whose metric can be written
locally in the form (\ref{metric of a pp-wave}).
\end{definition}

The advantage of Definition \ref{definition 1 of a pp-space}
is that it gives an explicit formula for the metric of a pp-space.
Its disadvantage is that it relies on a particular choice
of local coordinates in each coordinate patch.
We give now an alternative definition of a pp-space
which is much more geometrical.

\begin{definition}
\label{definition 2 of a pp-space}
A \,\emph{pp-space} is a Riemannian spacetime which
admits a nonvanishing parallel rank 1 spinor field.
\end{definition}

We use the term ``parallel''
to describe the situation when the covariant derivative of some tensor
or spinor field is identically zero.

It is known, see
Section 4 in \cite{Alekseevsky} or
Section 3.2.2 in \cite{Bryant}, that Definitions
\ref{definition 1 of a pp-space} and \ref{definition 2 of a pp-space}
are equivalent.

\begin{remark}
We do not assume that our spacetime admits a (global) spin structure,
cf. Section 11.6 in \cite{Nakahara}.
In fact, our only topological assumption is connectedness.
This does not prevent us from defining and parallel transporting spinors
locally.
\end{remark}

\begin{remark}
Whenever we deal with a parallel tensor or spinor field
we allow this field to be multivalued.
Multivaluedness may arise as a global phenomenon
when we combine the results of
local parallel transport along sections of a loop.
\end{remark}

\begin{remark}
In theoretical physics literature
the metric of a pp-wave is often characterised by the condition
that the spacetime admits a nonvanishing parallel
real null vector field, see, for example,
Section 21.5 in \cite{exact solutions}.
Here the authors assume implicitly the fulfillment of
the vacuum Einstein equation $Ric=0$.
It turns out,
see proof of Lemma \ref{lemma about parallel Ricci curvature},
that under the condition
$\nabla Ric=0$ the existence of a
nonvanishing parallel real null vector field
is equivalent to the existence of a
nonvanishing parallel rank 1 spinor field.
\end{remark}

Yet another way of characterising a pp-space is by
its restricted holonomy group $\operatorname{Hol}^0$.
Elementary calculations show that Definition
\ref{definition 2 of a pp-space} is equivalent to

\begin{definition}
\label{definition 3 of a pp-space}
A \,\emph{pp-space} is a Riemannian spacetime whose holonomy
$\operatorname{Hol}^0$
is, up to conjugation, a subgroup of the group
\begin{equation}
\label{definition of holonomy B2}
B^2:=\left\{
\left.
\left(
\begin{matrix}
\ 1\ &\ b\ \\
\ 0\ &\ 1\
\end{matrix}
\right)
\right|
\quad b\in\mathbb{C}
\right\}.
\end{equation}
\end{definition}

Here we use the standard identification of the proper
orthochroneous Lorentz group
with $\mathrm{SL}(2,\mathbb{C})$, see Example 5.57(c) in \cite{Nakahara}.
Our notation for subgroups of the proper Lorentz group
follows that of Section 10.122 of \cite{Besse};
note that some care is required because the text in \cite{Besse}
contains numerous mistakes.

It is interesting that the group
(\ref{definition of holonomy B2}) is, up to conjugation,
the unique nontrivial abelian Lie subgroup of $\mathrm{SL}(2,\mathbb{C})$.
In this statement ``nontrivial'' is understood as ``not 1-dimensional
and not a product of 1-dimensional subgroups'',
with dimension understood as real dimension.

Having stated the basic facts concerning pp-spaces, let us now return
to our analysis of the field equations
(\ref{eulerlagrangemetricLCexplicit}),
(\ref{eulerlagrangeconnectionLCexplicit}).
We claim that
pp-spaces with parallel Ricci curvature
are solutions of
(\ref{eulerlagrangemetricLCexplicit}),
(\ref{eulerlagrangeconnectionLCexplicit}).
The fact that such spacetimes satisfy
(\ref{eulerlagrangeconnectionLCexplicit}) is trivial, so we need
only to explain why they satisfy (\ref{eulerlagrangemetricLCexplicit}).
The Ricci curvature of a pp-space is, up to a scalar factor $s$,
the tensor square of a nonvanishing parallel
real null vector field $l$,
\begin{equation}
\label{type2equation1}
Ric_{\alpha\beta}=\,s\,l_\alpha l_\beta\,.
\end{equation}
In view of (\ref{type2equation1})
checking (\ref{eulerlagrangemetricLCexplicit})
reduces to checking
\begin{equation}
\label{type2equation2}
\mathcal{W}^{\kappa\lambda\mu\nu}l_\kappa l_\mu=0.
\end{equation}
In order to establish (\ref{type2equation2}) it is sufficient to
establish
\begin{equation}
\label{type2equation3}
\mathcal{W}^\kappa{}_{\lambda\mu\nu}\,l^\lambda=0.
\end{equation}
As the vector field $l$ is parallel we have
\begin{equation}
\label{type2equation4}
R^\kappa{}_{\lambda\mu\nu}\,l^\lambda=0.
\end{equation}
It remains to observe that formulae
(\ref{type2equation4}) and (\ref{type2equation1}) imply
(\ref{type2equation3}).

We call pp-spaces with parallel Ricci curvature
\emph{Riemannian solutions of type~2.}

In local coordinates
the function $s$ from (\ref{type2equation1})
is expressed via the function $f$ from (\ref{metric of a pp-wave})
as $s=c(f_{11}+f_{22})$ where
$f_{\alpha\beta}:=\frac{\partial^2f}{\partial x^\alpha x^\beta}$
and $c\ne0$ is some constant.
This observation gives us a simple algorithm for determining whether
the Ricci curvature of a pp-space is parallel.
Namely, the Ricci curvature of a
pp-space is parallel if and only if
$f_{11}+f_{22}=\operatorname{const}$,
and identically zero if and only if
$f_{11}+f_{22}=0$.
Note that in the latter case
the full (rank 4) curvature tensor $R$ is not necessarily zero
because it
is a linear function of the full Hessian
$(f_{\alpha\beta})_{\alpha,\beta=1}^2\,$,
and not only its trace.

\section{Riemannian solutions of type 3}
\label{type3}

Consider a Riemannian spacetime which has zero scalar curvature
and is locally a product of a pair of Einstein 2-manifolds.
(Of course, a 2-manifold is Einstein if and only if it has constant
curvature.)
Clearly, such a spacetime is a solution of
the field equation (\ref{eulerlagrangeconnectionLCexplicit}).
Straightforward calculations show that it is also a solution of
the field equation (\ref{eulerlagrangemetricLCexplicit}).

We call Riemannian spacetimes which have zero scalar curvature
and are locally a product of Einstein 2-manifolds
\emph{Riemannian solutions of type 3.}

The underlying reason why spacetimes described in this section
are indeed solutions of our field equations is as follows:
if we change the sign of the metric
of the Lorentzian 2-manifold
then the product becomes a
4-dimensional Einstein space.
Changing the sign of the metric
of the Lorentzian 2-manifold is equivalent to interchanging the
roles of the time and space coordinates.
This means that
a Riemannian solution of type 3 is
a Riemannian solution of type 1 with the wrong choice of
the time coordinate.
In other words, for all practical purposes
Riemannian solutions of type 3
are a special case
of Riemannian solutions of type 1.
We have to distinguish them only for the sake of mathematical bookkeeping.

\section{Uniqueness of Riemannian solutions}
\label{uniqueness}

The following uniqueness theorem is the main result of this paper.

\begin{theorem}
\label{uniqueness theorem}
Suppose that our coupling constants satisfy the inequalities
\begin{eqnarray}
\label{condition for uniqueness 1}
b_{911}&\ne& b_{922},
\\
\label{condition for uniqueness 2}
c_1+c_2&\ne&0,
\\
\label{condition for uniqueness 3}
c_1+c_3&\ne&0,
\\
\label{condition for uniqueness 4}
3c_1+c_2+2c_3&\ne&0.
\end{eqnarray}
Then solutions of types 1, 2 and 3
described in Sections \ref{type1}, \ref{type2} and \ref{type3}
respectively are the only Riemannian solutions of our field
equations
(\ref{eulerlagrangemetric}), (\ref{eulerlagrangeconnection}).
\end{theorem}

Here the $b$'s are the original coupling constants appearing in
formula (\ref{appendixb3}) whereas the $c$'s
are defined in accordance with formulae (\ref{formluae for cs}).

Note that conditions
(\ref{condition for uniqueness 2})--(\ref{condition for uniqueness 4})
appeared previously in \cite{pseudo}.
Namely, conditions
(\ref{condition for uniqueness 2}),
(\ref{condition for uniqueness 3})
coincide with condition (38) of \cite{pseudo},
whereas condition (\ref{condition for uniqueness 4})
is equivalent to the condition $c\ne-\frac13$
mentioned in the very end of Section 11 of \cite{pseudo}.
Thus, the new condition which enables us to
establish uniqueness is condition (\ref{condition for uniqueness 1}).
This matter will be discussed in greater detail in Section
\ref{Discussion}.

\begin{pf*}{Proof of Theorem \ref{uniqueness theorem}}
The crucial observation is that under the conditions
(\ref{condition for uniqueness 1}) and (\ref{condition for uniqueness 3})
the field equation
(\ref{eulerlagrangeconnectionLCexplicit})
is equivalent to
\begin{equation}
\label{covariantly constant ricci}
\nabla Ric=0.
\end{equation}
This fact is established by a sequence of elementary manipulations
with (\ref{eulerlagrangeconnectionLCexplicit}):
separate (\ref{eulerlagrangeconnectionLCexplicit})
into equations symmetric and antisymmetric in
the pair of indices $\kappa,\lambda\,$,
then contract
$\kappa$ with $\mu$ in the symmetric equation which
gives $\partial\mathcal{R}=0$, etc.
In performing these manipulations it is convenient
to express the constants $d_4,\ldots,d_7$
via the constants $c_1,c_2,c_3$ and $b_{911}-b_{922}$
in accordance with (\ref{formluae for ds via cs}).

Condition (\ref{covariantly constant ricci}) allows us to apply
the powerful
Lemma \ref{lemma about parallel Ricci curvature}.
The proof of Theorem \ref{uniqueness theorem} is therefore reduced
to the analysis of the situation when our spacetime
is locally a nontrivial product of Einstein manifolds,
with ``non\-trivial'' meaning that the spacetime itself is not Einstein.
We have to examine which
nontrivial products of Einstein manifolds satisfy
the field equation
(\ref{eulerlagrangemetricLCexplicit}),
and show that the only ones that do
are solutions of type 3 introduced in Section \ref{type3}.

The possible decompositions
into a nontrivial product are
3+1 and 2+2
where the numbers are the dimensions of Einstein manifolds.
Below we analyze each of these cases.
In doing this we use local coordinates which are a
concatenation of local coordinates on our Einstein manifolds;
consequently,
our metric and curvature have
block diagonal structure.
As usual, Greek letters in tensor indices run through
four possible values.
Note also that the 3+1 case actually splits into two subcases,
depending on whether the metric of the 3-manifold
is Euclidean or Lorentzian;
this distinction turns out to be unimportant because
the arguments presented below are insensitive to the signatures
of the metrics.

\textbf{Case 3+1.}
In this case
\begin{equation}
\label{decomposition of the metric}
g_{\mu\nu}=h_{\mu\nu}+k_{\mu\nu}
\end{equation}
where $h$ and $k$ are the metrics of the 3- and 1-manifolds
respectively, and
\[
R_{\kappa\lambda\mu\nu}
=\frac16\,
(h_{\kappa\mu}\,h_{\lambda\nu}-h_{\lambda\mu}\,h_{\kappa\nu})\,r
\]
where $r\ne0$ is the (constant) scalar curvature of the 3-manifold.
Straight\-forward calculations show that
in this case equation
(\ref{eulerlagrangemetricLCexplicit}) takes the form
\[
\frac{6d_2-d_3}{72}\,
(h^{\lambda\nu}-3k^{\lambda\nu})\,r^2=0.
\]
(Note the absence of the coefficient $d_1$ in this equation.
This is because in the 3+1 case Weyl curvature is zero.)
In view of
(\ref{formluae for ds via cs}) we have
$6d_2-d_3=c_1+c_2$, so under the condition
(\ref{condition for uniqueness 2}) the above equation cannot be satisfied.

\textbf{Case 2+2.}
In this case the metric is given by formula
(\ref{decomposition of the metric})
where $h$ and $k$ are the metrics of the two 2-manifolds, and
\[
R_{\kappa\lambda\mu\nu}
=\frac12\,
(h_{\kappa\mu}\,h_{\lambda\nu}-h_{\lambda\mu}\,h_{\kappa\nu})\,r
+\frac12\,
(k_{\kappa\mu}\,k_{\lambda\nu}-k_{\lambda\mu}\,k_{\kappa\nu})\,s
\]
where $r\ne s$ are the two corresponding (constant) scalar curvatures.
Straight\-forward calculations show that
in this case equation
(\ref{eulerlagrangemetricLCexplicit}) takes the form
\[
\frac{d_1+3d_2}{12}\,
(h^{\lambda\nu}-k^{\lambda\nu})\,(r^2-s^2)=0.
\]
In view of
(\ref{formluae for ds via cs}) we have
$d_1+3d_2=\frac12(3c_1+c_2+2c_3)$, so under the condition
(\ref{condition for uniqueness 4}) the above equation
is equivalent to $r+s=0$ which means that we are looking at
a solution of type 3, see Section \ref{type3}.
\qed
\end{pf*}

\section{The pseudoinstanton construction}
\label{The pseudoinstanton construction}

We are about to proceed to the study of non-Riemannian solutions
of our field equations
(\ref{eulerlagrangemetric}), (\ref{eulerlagrangeconnection}).
It would be unrealistic to expect to
find \emph{all} non-Riemannian solutions,
so we need a method for finding at least
\emph{some} non-Riemannian solutions. The following
construction provides such a method.

\begin{definition}
\label{definition of a pseudoinstanton}
We call a spacetime $\{M,g,\Gamma\}$
a \emph{pseudoinstanton}
if the connection is metric compatible
and curvature is irreducible and simple.
\end{definition}

Here irreducibility of curvature means that all irreducible pieces
but one are identically zero.
Simplicity means that the given irreducible subspace
is not isomorphic to any other irreducible subspace.
Metric compatibility means, as usual,
that $\nabla g\equiv0$.

The irreducible decomposition of curvature is described
in Appendix \ref{appendixa}.
It is easy to see that
there are only three possible types of pseudoinstantons:
\begin{itemize}
\item
\emph{scalar} pseudoinstanton
(all pieces of curvature
apart from the scalar piece $R^{(1)}$ are identically zero),
\item
\emph{pseudoscalar} pseudoinstanton
(all pieces of curvature
apart from the pseudo\-scalar piece $R_*^{(1)}$ are identically zero), and
\item
\emph{Weyl} pseudoinstanton
(all pieces of curvature
apart from the Weyl piece $R^{(10)}$ are identically zero).
\end{itemize}

\begin{theorem}
\label{pseudoinstanton theorem}
A pseudoinstanton is a solution
of the field equations
(\ref{eulerlagrangemetric}), (\ref{eulerlagrangeconnection}).
\end{theorem}

\begin{pf}
Put $R_{\text{pseudo}}:=R^{(1)}$
or $R_{\text{pseudo}}:=R_*^{(1)}$
or $R_{\text{pseudo}}:=R^{(10)}$,
depending on the type of our pseudoinstanton (see above).
Then
\[
q(\!R)=q(\!R_{\mathrm{pseudo}})+q(\!R-\!R_{\mathrm{pseudo}}).
\]
Note that
we used here the fact that the piece $R_{\mathrm{pseudo}}$ is simple:
if not, then we would have cross-over terms of the type
$R_{\mathrm{pseudo}}\times(\!R-\!R_{\mathrm{pseudo}})$.

When we start our variation from a spacetime with
$R-\!R_{\mathrm{pseudo}}\equiv0$
the resulting variation of $\,\int\!q(\!R-\!R_{\mathrm{pseudo}})\,$
is zero. Thus, the proof of Theorem
\ref{pseudoinstanton theorem} reduces to proving that
our pseudoinstanton is a stationary point of the action
$S_{\mathrm{pseudo}}:=\int\!q(\!R_{\mathrm{pseudo}})\,$.
But, according to Lemma \ref{lemma b1},
$\,q(\!R_{\mathrm{pseudo}})=c\,
(\!R_{\mathrm{pseudo}}\,,R_{\mathrm{pseudo}})_{\mathrm{YM}}\,$
where $c$ is some constant,
so the action $S_{\mathrm{pseudo}}$ is of the type studied in
\cite{pseudo} and the result follows from Theorem 2.1 of that paper.
\qed
\end{pf}

Further on we will be dealing only with the Weyl pseudoinstanton
as it is the most interesting of the three possible types
(the subspace of Weyl curvatures has dimension 10, whereas
the subspaces of scalar and pseudoscalar curvatures have dimension 1).
It is useful to rewrite Definition
\ref{definition of a pseudoinstanton} for this particular case.

\begin{definition}
\label{definition of a Weyl pseudoinstanton}
A \emph{Weyl pseudoinstanton} is a spacetime $\{M,g,\Gamma\}$
whose connection is metric compatible
and curvature purely Weyl.
\end{definition}

The advantage of Definition \ref{definition of a Weyl pseudoinstanton}
is that it can be used without knowledge of
the full irreducible decomposition of curvature
(material from Appendix \ref{appendixa}).
In particular, Weyl curvature can be understood
as curvature satisfying
$$
R_{\kappa\lambda\mu\nu}=R_{\mu\nu\kappa\lambda},
\qquad
Ric=0,
\qquad
\varepsilon^{\kappa\lambda\mu\nu}R_{\kappa\lambda\mu\nu}=0,
$$
which is the traditional definition. It is, of course,
equivalent to the definition given in Appendix \ref{appendixa}.

\section{Riemannian pseudoinstantons}
\label{Riemannian pseudoinstantons}

As mentioned in the beginning of Section
\ref{The pseudoinstanton construction},
the pseudoinstanton construction was developed primarily for
the purpose of finding non-Riemannian solutions to our
field equations
(\ref{eulerlagrangemetric}), (\ref{eulerlagrangeconnection}).
We already know the Riemannian solutions
(see Sections \ref{type1}--\ref{uniqueness} for details)
and these were found by means of a meticulous
straightforward analysis of the field equations.
Nevertheless, it is interesting to revisit the Riemannian
case using the pseudoinstanton technique.

A Riemannian spacetime is a Weyl pseudoinstanton if and only if
\begin{equation}
\label{vacuum Einstein equation}
Ric=0.
\end{equation}
Therefore, according to Theorem \ref{pseudoinstanton theorem},
Riemannian spacetimes satisfying the vacuum Einstein equation
(\ref{vacuum Einstein equation}) are solutions to our
field equations
(\ref{eulerlagrangemetric}), (\ref{eulerlagrangeconnection}).

The above argument demonstrates both the power and the limitations
of the pseudoinstanton technique.
This technique allowed us to obtain an important class of solutions
without having to write down explicitly the field equations.
On the other hand, it did not give us all the Riemannian solutions:
an Einstein space is not necessarily a pseudoinstanton,
and neither is a pp-space.

\section{Non-Riemannian pseudoinstantons}
\label{Non-Riemannian pseudoinstantons}

We know only one non-Riemannian solution,
and it is constructed as follows.
Let us define Minkowski space $\mathbb{M}^4$ as a
real 4-manifold equipped with
global coordinates $(x^0,x^1,x^2,x^3)$ and metric
$g_{\mu\nu}=\mathrm{diag}(+1,-1,-1,-1)$.
Let
\begin{equation}
\label{plane wave}
u(x)=w\,\mathrm{e}^{-\mathrm{i}l\cdot x}
\end{equation}
be a plane wave solution of the
equation $*\d u=\pm\mathrm{i}\d u$ in $\mathbb{M}^4$.
Define torsion
$T=\frac12\operatorname{Re}(u\otimes\d u)$,
and let $\Gamma$ be the corresponding metric compatible connection.
Then, as shown in \cite{pseudo},
the spacetime $\{\mathbb{M}^4,\Gamma\}$ is a Weyl pseudo\-instanton,
hence, by Theorem \ref{pseudoinstanton theorem},
a solution of our field equations
(\ref{eulerlagrangemetric}), (\ref{eulerlagrangeconnection}).

For the Yang--Mills case (\ref{YMq})
the ``torsion wave'' solution described above
was first obtained by Singh and Griffiths:
see last paragraph of Section 5 in \cite{griffiths3}
and put $k=0$, $N=\mathrm{e}^{-\mathrm{i}l\cdot x}$.
Our contribution is the observation that
this torsion wave remains a solution
for a general quadratic action (\ref{action})
and that this fact can be established
without having to write down explicitly the field equations.

The curvature of our spacetime $\{\mathbb{M}^4,\Gamma\}$ is
\begin{equation}
\label{curvature of torsion wave}
R=\operatorname{Re}(\d u\otimes\d u).
\end{equation}
The necessary and sufficient conditions for non-flatness are
$l\ne0$ and $w\not\in\operatorname{span}l$, and further
on we assume that these conditions are fulfilled.

We claim that
our spacetime $\{\mathbb{M}^4,\Gamma\}$ has holonomy $B^2$
(see formula (\ref{definition of holonomy B2})).
Indeed, it is easy to check that the tensor field $F:=l\wedge w$
is parallel.
But $*F=\pm\mathrm{i}F$ and $\det F=0$,
and the existence of a nonvanishing parallel tensor field
with such properties is equivalent to the statement
that the holonomy is a subgroup of the group $B^2$.
(In fact, a tensor $F$ with such properties is equivalent to a rank 1
spinor: this spinor is the ``square root'' of $F$, see
\cite{King and Vassiliev,kiev} for details.)
Finally, examination of formula
(\ref{curvature of torsion wave})
shows that the holonomy is at least 2-dimensional, hence it
coincides with $B^2$.
Comparing this result with Definitions
\ref{definition 2 of a pp-space}, \ref{definition 3 of a pp-space}
we conclude that our torsion wave
is a non-Riemannian analogue of a pp-space.

\section{Discussion}
\label{Discussion}


The main result of our paper is Theorem \ref{uniqueness theorem}
which is a uniqueness theorem for Riemannian solutions
of our metric-affine field equations
(\ref{eulerlagrangemetric}), (\ref{eulerlagrangeconnection}).

The problem of uniqueness of Riemannian solutions
in quadratic metric-affine gravity has a long history.
One of the first attempts at establishing uniqueness
was Fairchild's paper
\cite{fairchild1976}
in which the author considered equations
(\ref{eulerlagrangemetric}), (\ref{eulerlagrangeconnection})
in the Yang--Mills case (\ref{YMq}) and ``proved''
that Einstein spaces are the only solutions;
the mistake was acknowledged in \cite{fairchild1976erratum}.
In \cite{pseudo} the Yang--Mills action
was replaced by a more general quadratic action, the hope being
that this would break certain symmetries and lead to a uniqueness
result; unfortunately, the action in \cite{pseudo} still possessed
a substantial degree of symmetry and the uniqueness problem
was not resolved.
The basic difficulty with \cite{pseudo} was the fact that
the quadratic action was not general enough,
namely, it did not contain cross-over terms from pairs of
isomorphic pieces of curvature
whereas the isomorphic pieces themselves were chosen in an arbitrary
(traditional) fashion.
The reason why in the current paper we succeeded in obtaining
a uniqueness result for Riemannian solutions is that
the quadratic form $q$ appearing in (\ref{action})
is a \emph{most} general $\mathrm{O}(1,3)$-invariant
quadratic form on curvature.

An example of a quadratic form satisfying
the conditions of Theorem \ref{uniqueness theorem} is
\begin{equation}
\label{Ricciq}
q(\!R)=Ric_{\lambda\nu}\,Ric^{\lambda\nu}\,.
\end{equation}
In the representation (\ref{appendixb3})
the nonzero $b$'s for this quadratic form are
\[
b_1=1/4,\qquad
b_{611}=1,\qquad
b_{911}=1,
\]
hence the $c$'s defined in accordance
with formulae (\ref{formluae for cs}) are
\[
c_1=-1/2,\qquad
c_2=-3/2,\qquad
c_3=0.
\]
With these $b$'s and $c$'s
all four conditions of Theorem \ref{uniqueness theorem}
are satisfied.

Quadratic forms considered in \cite{pseudo} do not satisfy
the conditions of Theorem~\ref{uniqueness theorem} because
for such forms
condition (\ref{condition for uniqueness 1}) fails.
In particular, the Yang--Mills
quadratic form (\ref{YMq}) does not satisfy
the conditions of Theorem \ref{uniqueness theorem}.

The uniqueness Theorem \ref{uniqueness theorem}
still leaves us with the problem of providing a physical interpretation
for Riemannian solutions which are not Einstein spaces.
At the moment all we can say is that the set of such ``extra''
solutions is very meagre:
each is described locally by a real-valued function
$f(x^1,x^2,x^3)$ (see (\ref{metric of a pp-wave}))
satisfying, upon appropriate rescaling, the equation
$\frac{\partial^2f}{\partial(x^1)^2}+\frac{\partial^2f}{\partial(x^2)^2}=1$.


The second major result of our paper is the
non-Riemannian solution (``torsion wave'') described
in the beginning of Section
\ref{Non-Riemannian pseudoinstantons}.
It is tempting to view this torsion wave solution as a very basic
model for some elementary particle.
We nominate two candidates: the graviton and the neutrino.
Our (highly speculative) arguments go as follows.

Torsion is not an accepted physical observable
but curvature is, so we
base our interpretation on the analysis of the curvature
generated by our torsion wave.
Examination of the explicit formula (\ref{curvature of torsion wave})
indicates that it is more convenient to deal with the complexified
curvature $\d u\otimes\d u$; note also that complexification
is in line with the traditions of quantum mechanics.
Our complex curvature is polarized,
$\,{}^*(\d u\otimes\d u)=(\d u\otimes\d u)^*
=\pm\mathrm{i}(\d u\otimes\d u)\,$, and purely
Weyl, hence it is equivalent to a (symmetric) rank 4 spinor $\zeta$;
see subsection 1.2.3 in \cite{buchbinderandkuzenko}
or Appendix C in \cite{pseudo} for details.
A rank 4 spinor corresponds to a spin 2 particle,
and one naturally thinks of the graviton.

However, a closer examination reveals that our rank 4 spinor
has additional algebraic
structure: it is the 4th tensor power of a rank 1 spinor,
$\zeta=\xi\otimes\xi\otimes\xi\otimes\xi$.
Direct calculations \cite{King and Vassiliev,kiev}
show that the rank 1 spinor
field $\xi$ satisfies Weyl's equation,
which is the accepted mathematical model for the neutrino.

It is worth pointing out that we actually have
two nonvanishing
rank 1 spinor fields associated with our torsion wave solution:
one is the spinor field $\xi$ from the previous paragraph,
and the other is the parallel spinor field $\xi^\parallel$.
The latter exists because our torsion wave solution has
holonomy $B^2$, see end of Section~\ref{Non-Riemannian pseudoinstantons}.
The two spinor fields differ by a scalar factor, namely,
\begin{equation}
\label{almost parallel}
\xi=c\,\mathrm{e}^{-\frac{\mathrm{i}}2\int l\cdot\d x}\,\xi^\parallel
\end{equation}
where $c\ne0$ is some constant and
$l$ is the nonvanishing parallel real null (co)vector field
from formula (\ref{plane wave}).
In Alekseevsky's terminology \cite{Alekseevsky},
$\xi$ and $\xi^\parallel$ are \emph{conformally equivalent}
and $\xi$ is \emph{almost parallel.}
Given a standard choice of coordinates and Pauli matrices
in Minkowski space,
it is easy to see that the components of $\xi^\parallel$ are constant,
so formula (\ref{almost parallel}) implies
$\nabla\xi=\partial\xi$
(covariant derivative coincides with partial derivative).
Thus, in our non-Riemannian spacetime
the covariant derivative of the spinor field $\xi$
is the same as in flat Minkowski space.

It is also worth pointing out that in our non-Riemannian spacetime
Weyl's equation is the same as in flat Minkowski space:
this is a consequence of the fact that
our torsion is purely tensor,
see \cite{ryder} and  \cite{pseudo} for details.

The above arguments indicate that observation of our torsion wave solution
may lead the observer to believe that they are in a Riemannian spacetime.

\appendix

\section{Irreducible decomposition of curvature}
\label{appendixa}

A curvature generated by a
general affine connection has only one (anti)sym\-metry, namely,
\begin{equation}
\label{appendixa1}
R^\kappa{}_{\lambda\mu\nu}=-R^\kappa{}_{\lambda\nu\mu}\,.
\end{equation}
For a fixed $x\in M$
we denote by $\mathbf{R}$ the 96-dimensional vector space
of real rank 4 tensors $R^\kappa{}_{\lambda\mu\nu}$
satisfying condition (\ref{appendixa1}).

Let $g$ be the Lorentzian metric at the point $x\in M$ and let
$\mathrm{O}(1,3)$ be the corresponding full Lorentz group, i.e. the group
of linear transformations of coordinates in the tangent space $T_xM$
which preserve the metric.
It is known, see Appendix B.4 from \cite{hehlreview},
that the vector space $\mathbf{R}$ decomposes into a direct sum
of 11 subspaces which are invariant and irreducible
under the action of $\mathrm{O}(1,3)$.
These subspaces are listed in Table A.1.
\begin{table}
\caption{List of irreducible subspaces}
\begin{tabular}{c|c|c}
Dimension&Number of subspaces&Notation for subspaces\\
\hline
1&2&$\mathbf{R}^{(1)}$, $\ \mathbf{R}_*^{(1)}$\\
\hline
6&3&$\mathbf{R}^{(6,l)}$, $\ l=1,2,3$\\
\hline
9&4&$\mathbf{R}^{(9,l)}$, $\ \mathbf{R}_*^{(9,l)}$, $\ l=1,2$\\
\hline
10&1&$\mathbf{R}^{(10)}$\\
\hline
30&1&$\mathbf{R}^{(30)}$
\end{tabular}
\end{table}
Note that our notation
differs from that of \cite{hehlreview}:
we want to emphasize
the fact that there are 3 groups of isomorphic subspaces, namely,
\begin{equation}
\label{appendixa2}
\{\mathbf{R}^{(6,l)},\ l=1,2,3\},\qquad
\{\mathbf{R}^{(9,l)},\ l=1,2\},\qquad
\{\mathbf{R}_*^{(9,l)},\ l=1,2\}.
\end{equation}
Two subspaces are said to be isomorphic
is there is a linear bijection between them
which commutes with the action of $\mathrm{O}(1,3)$.

In order to give an explicit description of irreducible subspaces
of curvature we introduce the following conventions.
We lower and raise tensor indices using the
metric, and we also denote
\begin{equation}
\label{appendixa3}
(R^*)_{\kappa\lambda\mu\nu}:=
\frac12\,\sqrt{|\det g|}
\ R_{\kappa\lambda\mu'\nu'}\,
\varepsilon^{\mu'\nu'}{}_{\mu\nu}\,.
\end{equation}
The map $R\to R^*$ is an endomorphism in $\mathbf{R}$
which we call the \emph{right Hodge star}.
Note that as we are working in the real Lorentzian setting the
Hodge star has \emph{no} eigenvalues.

The explicit description of irreducible subspaces of dimension $<10$
is given in Table A.2.
\begin{table}
\caption
{Explicit description of irreducible subspaces of dimension $<10$}
\begin{tabular}{c|c}
Subspace&Formula for curvature $R$\\
\hline
$\mathbf{R}^{(1)}$&$R_{\kappa\lambda\mu\nu}=a_1
(g_{\kappa\mu}g_{\lambda\nu}-g_{\kappa\nu}g_{\lambda\mu})
\mathcal{R}$\\
\hline
$\mathbf{R}_*^{(1)}$&$(\!R^*\!)_{\kappa\lambda\mu\nu}=a^*_1
(g_{\kappa\mu}g_{\lambda\nu}-g_{\kappa\nu}g_{\lambda\mu})
\mathcal{R}_*$\\
\hline
$\mathbf{R}^{(6,l)}$&$R_{\kappa\lambda\mu\nu}=
a_{6l1}
(g_{\kappa\mu}\mathcal{A}^{(l)}{}_{\lambda\nu}
-g_{\kappa\nu}\mathcal{A}^{(l)}{}_{\lambda\mu})
+a_{6l2}
(g_{\lambda\mu}\mathcal{A}^{(l)}{}_{\kappa\nu}
-g_{\lambda\nu}\mathcal{A}^{(l)}{}_{\kappa\mu})$\\
&\hfill$+a_{6l3}
g_{\kappa\lambda}\mathcal{A}^{(l)}{}_{\mu\nu}$\\
\hline
$\mathbf{R}^{(9,l)}$&$R_{\kappa\lambda\mu\nu}=
a_{9l1}
(g_{\kappa\mu}\mathcal{S}^{(l)}{}_{\lambda\nu}
-g_{\kappa\nu}\mathcal{S}^{(l)}{}_{\lambda\mu})
+a_{9l2}
(g_{\lambda\mu}\mathcal{S}^{(l)}{}_{\kappa\nu}
-g_{\lambda\nu}\mathcal{S}^{(l)}{}_{\kappa\mu})$\\
\hline
$\mathbf{R}_*^{(9,l)}$&$(\!R^*\!)_{\kappa\lambda\mu\nu}=
a^*_{9l1}
(g_{\kappa\mu}\mathcal{S}_*^{(l)}{}_{\lambda\nu}
-g_{\kappa\nu}\mathcal{S}_*^{(l)}{}_{\lambda\mu})
+a^*_{9l2}
(g_{\lambda\mu}\mathcal{S}_*^{(l)}{}_{\kappa\nu}
-g_{\lambda\nu}\mathcal{S}_*^{(l)}{}_{\kappa\mu})$
\end{tabular}
\end{table}
Here
$\mathcal{R}$, $\mathcal{R}_*$ are arbitrary scalars,
$\mathcal{A}^{(l)}$ are arbitrary
rank 2 antisymmetric tensors, and
$\mathcal{S}^{(l)}$, $\mathcal{S}_*^{(l)}$ are arbitrary
rank 2 symmetric trace-free tensors,
with ``arbitrary'' meaning that the quantity in question
spans its vector space.
The $a$'s in Table A.2 are some fixed real constants,
the only condition being that
$a_1$,
$a^*_1$,
$\det\,(a_{6lm})_{l,m=1}^3\,$,
$\det\,(a_{9lm})_{l,m=1}^2\,$ and
$\det\,(a^*_{9lm})_{l,m=1}^2\,$
are nonzero.
The freedom in choosing irreducible subspaces of
dimension 6 and 9 is due to the fact that we have groups
of isomorphic subspaces (\ref{appendixa2}).

It is convenient to choose the following $a$'s:
\begin{multline}
\label{appendixa4}
a_1=a^*_1=\frac1{12},\\
(a_{6lm})
=\left(
\begin{array}{ccc}
\frac5{12}&-\frac1{12}&-\frac16\\
-\frac1{12}&\frac5{12}&-\frac16\\
-\frac1{12}&-\frac1{12}&\frac13
\end{array}
\right),\quad
(a_{9lm})=(a^*_{9lm})
=\left(
\begin{array}{cc}
\frac38&-\frac18\\
-\frac18&\frac38
\end{array}
\right).
\end{multline}
Then the lower rank tensors
$\mathcal{R}$, $\mathcal{R}_*$,
$\mathcal{A}^{(l)}$,
$\mathcal{S}^{(l)}$, $\mathcal{S}_*^{(l)}$
appearing in Table A.2
are expressed via the full (rank 4) curvature tensor $R\,$
according to the following simple formulae:
\begin{multline}
\label{appendixa5}
\mathcal{R}:=R^{\kappa\lambda}{}_{\kappa\lambda},
\\
Ric^{(1)}{}_{\lambda\nu}:=R^\kappa{}_{\lambda\kappa\nu},
\qquad
Ric^{(2)}{}_{\kappa\nu}:=R_\kappa{}^\lambda{}_{\lambda\nu},
\\
\mathcal{R}ic^{(1)}:=Ric^{(1)}-\frac14\mathcal{R}g,
\qquad
\mathcal{R}ic^{(2)}:=Ric^{(2)}+\frac14\mathcal{R}g,
\\
\mathcal{S}^{(l)}{}_{\mu\nu}:=\frac
{\mathcal{R}ic^{(l)}{}_{\mu\nu}+\mathcal{R}ic^{(l)}{}_{\nu\mu}}2,
\quad
\mathcal{A}^{(l)}{}_{\mu\nu}:=\frac
{\mathcal{R}ic^{(l)}{}_{\mu\nu}-\mathcal{R}ic^{(l)}{}_{\nu\mu}}2,
\quad l=1,2,
\\
\mathcal{A}^{(3)}{}_{\mu\nu}:=R^\kappa{}_{\kappa\mu\nu},
\end{multline}
and
\begin{multline}
\label{appendixa6}
\mathcal{R}_*:=(R^*)^{\kappa\lambda}{}_{\kappa\lambda},
\\
Ric_*^{(1)}{}_{\lambda\nu}:=(R^*)^\kappa{}_{\lambda\kappa\nu},
\qquad
Ric_*^{(2)}{}_{\kappa\nu}:=(R^*)_\kappa{}^\lambda{}_{\lambda\nu},
\\
\mathcal{R}ic_*^{(1)}:=Ric_*^{(1)}-\frac14\mathcal{R}_*g,
\qquad
\mathcal{R}ic_*^{(2)}:=Ric_*^{(2)}+\frac14\mathcal{R}_*g,
\\
\mathcal{S}_*^{(l)}{}_{\mu\nu}:=\frac
{\mathcal{R}ic_*^{(l)}{}_{\mu\nu}+\mathcal{R}ic_*^{(l)}{}_{\nu\mu}}2,
\quad
\mathcal{A}_*^{(l)}{}_{\mu\nu}:=\frac
{\mathcal{R}ic_*^{(l)}{}_{\mu\nu}-\mathcal{R}ic_*^{(l)}{}_{\nu\mu}}2,
\quad l=1,2,
\\
\mathcal{A}_*^{(3)}{}_{\mu\nu}:=(R^*)^\kappa{}_{\kappa\mu\nu}.
\end{multline}
Note that the tensors $\mathcal{A}_*^{(l)}$ are not used
in Table A.2. This is not surprising as the tensors
$\mathcal{A}^{(l)}$ and $\mathcal{A}_*^{(l)}$ are not independent:
the $\mathcal{A}^{(l)}$ are linear combinations of the
Hodge duals of $\mathcal{A}_*^{(l)}$ and vice versa.

All calculations in the main text of the
paper use the (\ref{appendixa4}) choice of $a$'s.

Finally, let us give an explicit description of the
10- and 30-dimensional irreducible subspaces.
$\,\mathbf{R}^{(10)}$ is the
subspace of curvatures $R\,$ such that
\begin{equation}
\label{appendixa7}
R^\kappa{}_{\lambda\kappa\nu}
=(R^*)^\kappa{}_{\lambda\kappa\nu}=0,\qquad
R_\kappa{}^\lambda{}_{\lambda\nu}
=(R^*)_\kappa{}^\lambda{}_{\lambda\nu}=0,\qquad
R^\kappa{}_{\kappa\mu\nu}=0
\end{equation}
(all possible traces are zero)
and $R_{\kappa\lambda\mu\nu}=-R_{\lambda\kappa\mu\nu}$.
$\,\mathbf{R}^{(30)}$ is the
subspace of curvatures $R\,$ satisfying (\ref{appendixa7})
and $R_{\kappa\lambda\mu\nu}=R_{\lambda\kappa\mu\nu}$.

Given a decomposition
\[
\mathbf{R}
=\mathbf{R}^{(1)}
\oplus\mathbf{R}_*^{(1)}
\oplus_{l=1}^3\mathbf{R}^{(6,l)}
\oplus_{l=1}^2\mathbf{R}^{(9,l)}
\oplus_{l=1}^2\mathbf{R}_*^{(9,l)}
\oplus\mathbf{R}^{(10)}
\oplus\mathbf{R}^{(30)}
\]
any $R\in\mathbf{R}$ can be uniquely written as
\[
R
=R^{(1)}
+R_*^{(1)}
+\sum_{l=1}^3R^{(6,l)}
+\sum_{l=1}^2R^{(9,l)}
+\sum_{l=1}^2R_*^{(9,l)}
+R^{(10)}
+R^{(30)}
\]
where the $R$'s in the RHS are from the corresponding irreducible
subspaces.
We will call these $R$'s the
\emph{irreducible pieces of curvature.}
We will call the irreducible pieces
$R^{(1)}$, $R_*^{(1)}$, $R^{(10)}$, $R^{(30)}$ \emph{simple}
because their subspaces are not isomorphic to any other subspaces.

\begin{remark}
\label{appendixaremark1}
``Starred'' and ``unstarred'' subspaces of same dimension are not
isomorphic under the action of the group $\mathrm{O}(1,3)$
because the Hodge star is a linear map which depends on
the choice of the element of the group.
This dependence is encoded in the normalisation of the totally
antisymmetric quantity:
$\varepsilon_{0123}=+1$ or $\varepsilon_{0123}=-1$
depending on whether the orientation of the coordinate
system is positive or negative.
\end{remark}

\begin{remark}
\label{appendixaremark2}
The global definition of the Hodge star requires the orientability
of our manifold $M$.
However, for the purpose of decomposing curvatures orientability
is not needed: any abstract vector subspace is preserved under inversion
($\mathrm{vector}\mapsto-\mathrm{vector}$),
so when writing explicit formulae for subspaces
it does not matter whether
$\varepsilon_{0123}=+1$ or $\varepsilon_{0123}=-1$.
The delicate features of the Hodge star come to light only when
we examine the relationship between \emph{pairs} of different subspaces,
see Remark \ref{appendixaremark1}.
\end{remark}

\begin{remark}
\label{appendixaremark3}
If we complexify our problem then our 11 subspaces will still
remain irreducible under the action of $\mathrm{O}(1,3)$.
In order to justify this claim we argue as follows.
Replace the full Lorentz group $\mathrm{O}(1,3)$ by the proper
orthochroneous Lorentz
group $\mathrm{SO}(1,3)^\uparrow$. Then we have the standard algorithm
(see, for example, Section 1.2 in \cite{buchbinderandkuzenko})
for finding irreducible subspaces in terms of spinors.
Applying this algorithm we see that the complexified subspaces
$\mathbf{R}^{(6,l)}$, $\mathbf{R}^{(10)}$ and $\mathbf{R}^{(30)}$
split into eigenspaces of the right Hodge star (\ref{appendixa3}),
and these ``halves'' are the only proper
$\mathrm{SO}(1,3)^\uparrow$-invariant subspaces
of the original subspaces.
However, the ``halves'' are not invariant under change of orientation.
\end{remark}

\section{Quadratic forms on curvature}
\label{appendixb}

Let us define an inner product on rank 2 tensors
\begin{equation}
\label{appendixb1}
(K,L):=K_{\mu\nu}\,L^{\mu\nu}\,,
\end{equation}
and a Yang--Mills inner product on curvatures
\begin{equation}
\label{appendixb2}
(R,Q)_{\mathrm{YM}}:=R^\kappa{}_{\lambda\mu\nu}\,
Q^\lambda{}_\kappa{}^{\mu\nu}\,.
\end{equation}

\begin{lemma}
\label{lemma b1}
Let $q:\mathbf{R}\to\mathbb{R}$ be an
$\mathrm{O}(1,3)$-invariant quadratic form on curvature. Then
\begin{multline}
\label{appendixb3}
q(\!R)=b_1\mathcal{R}^2+b^*_1\mathcal{R}_*^2
\\
+\sum_{l,m=1}^3 b_{6lm}(\mathcal{A}^{(l)},\mathcal{A}^{(m)})
+\sum_{l,m=1}^2 b_{9lm}(\mathcal{S}^{(l)},\mathcal{S}^{(m)})
+\sum_{l,m=1}^2 b^*_{9lm}(\mathcal{S}_*^{(l)},\mathcal{S}_*^{(m)})
\\
+b_{10}(R^{(10)},R^{(10)})_{\mathrm{YM}}
+b_{30}(R^{(30)},R^{(30)})_{\mathrm{YM}}
\end{multline}
with some real constants
$b_1$, $b^*_1$,
$b_{6lm}=b_{6ml}$,
$b_{9lm}=b_{9ml}$,
$b^*_{9lm}=b^*_{9ml}$,
$b_{10}$, $b_{30}$.
Here
$\mathcal{R}$, $\mathcal{R}_*$,
$\mathcal{A}^{(l)}$,
$\mathcal{S}^{(l)}$, $\mathcal{S}_*^{(l)}$,
$R^{(10)}$, $R^{(30)}$
are tensors defined in Appendix \ref{appendixa}.
\end{lemma}

\begin{pf}
Let us equip each of the 11 irreducible subspaces of curvature
(see Table A.1) with an $\mathrm{O}(1,3)$-invariant
non-degenerate inner product.
For 1-dimensional subspaces we employ the usual multiplication
of scalars, and for 6- and 9-dimensional subspaces we employ
(\ref{appendixb1}); here scalars and rank 2 tensors are related to
irreducible pieces of curvature in accordance with Table A.2.
Note that these inner products are well defined even if the manifold $M$
is non-orientable: the fact that in a ``starred'' subspace our scalar
or rank 2 tensor may be defined up to sign has no bearing on the
inner product because both entries in the inner product would
simultaneously retain or change sign
upon continuation over a loop in $M$.
See also Remark \ref{appendixaremark2}.
Our inner products on 1-, 6- and 9-dimensional subspaces are clearly
non-degenerate.

For 10- and 30-dimensional subspaces we employ
the inner product (\ref{appendixb2}). It is not \emph{a priori}
clear that this inner product is non-degenerate on these subspaces.
We establish non-degeneracy as follows.
The inner product (\ref{appendixb2}) is clearly non-degenerate on the
whole vector space $\mathbf{R}$. It is easy to check that any pair
of non-isomorphic subspaces is orthogonal with respect to the inner
product (\ref{appendixb2}), so $\mathbf{R}$ decomposes into a direct
sum of 7 orthogonal subspaces
\[
\mathbf{R}^{(1)},\quad
\mathbf{R}_*^{(1)},\quad
\oplus_{l=1}^3\mathbf{R}^{(6,l)},\quad
\oplus_{l=1}^2\mathbf{R}^{(9,l)},\quad
\oplus_{l=1}^2\mathbf{R}_*^{(9,l)},\quad
\mathbf{R}^{(10)},\quad
\mathbf{R}^{(30)}.
\]
Hence, the inner product (\ref{appendixb2}) is non-degenerate
on each of these 7 subspaces.
In particular, it is non-degenerate on
$\mathbf{R}^{(10)}$ and $\mathbf{R}^{(30)}$.

Further on in the proof
we deal with the bilinear form
$b:\mathbf{R}\times\mathbf{R}\to\mathbb{R}$
associated with the quadratic form $q$, i.e. $q(\!R)=b(\!R\,,\!R)$.

Let $V$ and $W$ be irreducible subspaces of $\mathbf{R}$
and let
$(\,\cdot\,,\,\cdot\,)_V$ and $(\,\cdot\,,\,\cdot\,)_W$
be their $\mathrm{O}(1,3)$-invariant non-degenerate inner products.
Here $V$ and $W$ are not necessarily distinct.
Consider the $\mathrm{O}(1,3)$-invariant
bilinear form $b_{VW}:=\left.b\right|_{V\times W}$.
Then there is a unique linear map $B_{VW}:V\to W$ such that
\[
(B_{VW}v,w)_W=b_{VW}(v,w),\qquad\forall v\in V,\quad\forall w\in W,
\]
and this map commutes with the action of $\mathrm{O}(1,3)$.
By Schur's lemma $B_{VW}$ is either zero or a bijection,
in which case $V$ and $W$ are isomorphic.
Thus, only pairs of isomorphic irreducible subspaces can give nonzero
contributions to the bilinear form $b$.

The proof of Lemma \ref{lemma b1} has been reduced to proving the
following fact:
if $V$ is an irreducible subspace of $\mathbf{R}$ and
$B_V:V\to V$ is a linear operator which
commutes with the action of $\mathrm{O}(1,3)$
then $B_V$ is a multiple of the identity map.
In order to prove this fact we complexify our problem,
noting that
by Remark \ref{appendixaremark3} this does not affect the
irreducibility of $V$.
After complexification
the fact we are proving becomes a special case of a
well known abstract result.
\qed
\end{pf}

\section{Spacetimes with parallel Ricci curvature}
\label{appendixc}

The purpose of this appendix is to state and prove the following

\begin{lemma}
\label{lemma about parallel Ricci curvature}
A Riemannian spacetime has parallel Ricci curvature if and only if
\begin{itemize}
\item[\emph{(a)}]
it is locally a product of Einstein manifolds, or
\item[\emph{(b)}]
it is a pp-space with parallel Ricci curvature
(see Section \ref{type2}).
\end{itemize}
\end{lemma}

Before proceeding to the proof of Lemma
\ref{lemma about parallel Ricci curvature}
let us recall that throughout this paper
our spacetime is assumed to be
4-dimensional, real, connected and equipped
with \emph{Lorentzian} metric; see also Remark \ref{remark on riemannian}
on the meaning of ``Riemannian''.
All these assumptions are important in Lemma
\ref{lemma about parallel Ricci curvature}.

The notion of an Einstein manifold is understood as in
Definition 1.95 of \cite{Besse}: a real manifold of arbitrary dimension
equipped with a pseudo-Euclidean metric and
a Levi-Civita connection, and such that the Ricci tensor is
proportional to the metric with a \emph{constant} proportionality
factor.

Note that
Lemma \ref{lemma about parallel Ricci curvature}
has a well-known
Euclidean analogue.
Namely, in the Euclidean case Ricci curvature is parallel
if and only if the mani\-fold
is locally a product of  Einstein manifolds;
see Theorem 1.100 and Section 16.A in \cite{Besse}.

\begin{pf*}{Proof of Lemma \ref{lemma about parallel Ricci curvature}}
The fact that assertion (a) or (b) implies
(\ref{covariantly constant ricci}) is obvious,
so we only need to prove the converse statement.

It is known, see \cite{Alekseevsky} or Section 10.119 in \cite{Besse},
that our spacetime $(M,g)$ is, at least locally, a product of
pseudo-Euclidean manifolds
$(M_j,g_j)$, $j=1,\dots,k$, whose holonomies
are weakly irreducible.
Here ``weak irreducibility'' means that
the only non-degenerate (with respect to the metric) invariant subspaces of
the tangent space are $\{0\}$ and the tangent space itself.
Condition (\ref{covariantly constant ricci})
implies
\begin{equation}
\label{covariantly constant ricci j}
\nabla Ric_j=0,
\end{equation}
$j=1,\dots,k$, where $Ric_j$ is the
Ricci curvature of $(M_j,g_j)$.

Let us examine a given manifold $(M_j,g_j)$.
If $\dim M_j=1$ then $(M_j,g_j)$ is clearly Einstein.
If $\dim M_j=2$ then
(\ref{covariantly constant ricci j}) implies that
$(M_j,g_j)$ is Einstein.
If $\dim M_j=3$ or $\dim M_j=4$ then
$(M_j,g_j)$ may not be Einstein, in which case,
in view of (\ref{covariantly constant ricci j}), it admits
a nonvanishing parallel symmetric
rank 2 trace-free tensor field.
But all such manifolds have been classified, see Table~2 in
\cite{Alekseevsky}.
Analysis of the latter shows that if our spacetime is
not a product of Einstein manifolds then we have
one of the following three cases:
\begin{eqnarray}
\label{decomposition case 1}
\operatorname{Hol}^0&=&A^1\times\{1\},
\\
\label{decomposition case 2}
\operatorname{Hol}^0&=&B^2,
\\
\label{decomposition case 3}
\operatorname{Hol}^0&=&B^3_{\mathrm{i}}.
\end{eqnarray}
Here
\[
A^1:=\left\{
\left.
\left(
\begin{matrix}
\ 1\ &\ b\ \\
\ 0\ &\ 1\
\end{matrix}
\right)
\right|
\quad b\in\mathbb{R}
\right\},
\quad
B^3_{\mathrm{i}}:=\left\{
\left.
\left(
\begin{array}{cc}
\ a\ &\ b\ \\
\ 0\ &\ a^{-1}\
\end{array}
\right)
\right|
\quad a,b\in\mathbb{C},\quad|a|=1
\right\},
\]
and $B^2$ is defined in accordance with
(\ref{definition of holonomy B2}); note that
we continue using the notation from
Section 10.122 of \cite{Besse}.
Cases (\ref{decomposition case 1}) and (\ref{decomposition case 2})
correspond to pp-spaces (see Definition \ref{definition 3 of a pp-space}),
whereas (\ref{decomposition case 3}) does not.
It remains to
show that the case (\ref{decomposition case 3}) is impossible.

In the remainder of the proof we assume that we have
(\ref{decomposition case 3}).
We will show that this leads to a contradiction.

Condition (\ref{decomposition case 3}) implies
the existence of a nonvanishing parallel
real null vector field $l$.
This condition also restricts
the possible structure of the full (rank~4) curvature tensor $R$.
To understand the latter let us fix an arbitrary point $x\in M$
and choose a pair of real vectors $v_1$, $v_2$ such that
$l\cdot v_1=l\cdot v_2=v_1\cdot v_2=0$ and
$v_1\cdot v_1=v_2\cdot v_2=-1$
where the dot denotes the standard inner product on $T_xM$.
Put $A_j:=l\wedge v_j$, $j=1,2$, $A_3:=v_1\wedge v_2$.
It is easy to see that $\{A_1,A_2,A_3\}$ is a basis
for $\mathfrak{b}^3_{\mathrm{i}}$,
the Lie algebra of the group $B^3_{\mathrm{i}}$.
Also, $\{A_1,A_2\}$ is a basis for $\mathfrak{b}^2$,
the Lie algebra of the group $B^2$.
Condition (\ref{decomposition case 3}) implies that at
the point $x$ the curvature tensor has the structure
\begin{equation}
\label{structure of R}
R=\sum_{j,k=1}^3c_{jk}\,A_j\otimes A_k
\end{equation}
where $c_{jk}=c_{kj}$ are some real numbers.

Further on we denote by $u^2:=u\otimes u$
the tensor square of a vector,
and by $u\vee v:=u\otimes v+v\otimes u$ the symmetric product of
a pair of vectors.

We have (\ref{covariantly constant ricci}) and, therefore,
$\nabla\mathcal{R}ic=0$.
According to Table~2 in \cite{Alekseevsky}, under the condition
(\ref{decomposition case 3}) the only (up to rescaling)
nonvanishing parallel symmetric trace-free
rank 2 tensor field is $l^2$, hence $\mathcal{R}ic$ is a multiple of $l^2$.
But formula (\ref{structure of R}) implies
\[
\mathcal{R}ic=-(c_{11}+c_{22})l^2
+c_{13}\,l\vee v_2-c_{23}\,l\vee v_1
-c_{33}\left(\frac12g+v_1^2+v_2^2\right),
\]
so $\mathcal{R}ic$ is a multiple of $l^2$ if and only if
$c_{13}=c_{23}=c_{33}=0$.
Formula (\ref{structure of R}) now becomes
\begin{equation}
\label{structure of R, improved}
R=\sum_{j,k=1}^2c_{jk}\,(l\wedge v_j)\otimes(l\wedge v_k)\,.
\end{equation}

Denote $L:=\operatorname{span}l\subset T_xM$,
$L^\perp:=\{u|\ u\perp l\}\subset T_xM$,
and let $R_{\mu\nu}:T_xM\to T_xM$ be the linear operator
defined by
$(R_{\mu\nu}\,u)^\kappa:=R^\kappa{}_{\lambda\mu\nu}\,u^\lambda$.
Inspection of formula (\ref{structure of R, improved}) shows that
\begin{equation}
\label{fundamental property of R}
R_{\mu\nu}(L^\perp)\subset L.
\end{equation}
A convenient way of interpreting this result is to think of a
connection on $L^\perp/L\ $:
then (\ref{fundamental property of R})
is the statement that the curvature of such a
connection is zero.
(The connection on $L^\perp/L$ is, in fact,
equivalent to a $\mathrm{U}(1)$-connection.)

Let us now fix a point $x_0\in M$.
Put $l_0:=\left.l\right|_{x=x_0}$,
$L_0:=\left.L\right|_{x=x_0}$,
$L_0^\perp:=\left.L^\perp\right|_{x=x_0}$.
Let $\Lambda$ be an arbitrary loop based at $x_0$
which is homotopic to the constant loop at $x_0$.
Denote by $h_\Lambda:T_{x_0}M\to T_{x_0}M$
the linear operator describing the result of parallel transport of
a vector along this loop.
As the vector field $l$ is parallel we have
\begin{equation}
\label{holonomy formula 1}
(h_\Lambda-\operatorname{id})(l_0)=0.
\end{equation}
In view of (\ref{fundamental property of R}) we also have
\begin{equation}
\label{holonomy formula 2}
(h_\Lambda-\operatorname{id})(L_0^\perp)\subset L_0.
\end{equation}
It is easy to see that properties
(\ref{holonomy formula 1}) and (\ref{holonomy formula 2}) imply
$\operatorname{Hol}^0\le B^2$,
which contradicts (\ref{decomposition case 3}).
Thus, the case (\ref{decomposition case 3}) is impossible.
\qed
\end{pf*}

\begin{ack}

The author is indebted to
D.V.~Alekseevsky and J.B.~Griffiths for helpful advice.
Special thanks go to F.E.~Burstall who suggested an alternative,
fully self-contained proof
of Lemma \ref{lemma about parallel Ricci curvature}.

\end{ack}


\begin{thebibliography}{19}




\bibitem{ryder}
M. Adak, T. Dereli, L.H. Ryder,
Neutrino oscillations induced by spacetime torsion,
Class. Quantum Grav. 18 (2001) 1503--1512.

\bibitem{Alekseevsky}
D.V. Alekseevsky,
Holonomy groups and recurrent tensor fields in Lorentzian spaces,
in: Problems of the Theory of Gravitation and Elementary Particles
issue 5 (ed. K.P. Stanjukovich),
Atomizdat, Moscow, 1974, pp. 5--17. In Russian.


\bibitem{Besse}
A.L. Besse,
Einstein Manifolds,
Springer-Verlag, Berlin, 1987.

\bibitem{Bryant}
R.L. Bryant,
Pseudo-Riemannian metrics with parallel spinor fields
and vanishing Ricci tensor,
in: Global Analysis and Harmonic Analysis (Marseille-Luminy, 1999),
S\'emin. Congr., 4, Soc. Math. France, Paris, 2000,
pp. 53–-94.

\bibitem{buchbinderandkuzenko}
I.L. Buchbinder, S.M. Kuzenko,
Ideas and Methods of Supersymmetry and Supergravity,
Institute of Physics Publishing, Bristol, 1998.


\bibitem{fairchild1976}
E.E. Fairchild (Jr.),
Gauge theory of gravitation,
Phys. Rev. D 14 (1976) 384--391.

\bibitem{fairchild1976erratum}
E.E. Fairchild (Jr.),
Erratum: Gauge theory of gravitation,
Phys. Rev. D 14 (1976) 2833.

\bibitem{hehlreview}
F.W. Hehl, J.D. McCrea, E.W. Mielke, Y. Ne'eman,
Metric-affine gauge theory of gravity: field equations,
Noether identities, world spinors, and breaking of dilation invariance,
Phys. Rep. 258 (1995) 1--171.

\bibitem{King and Vassiliev}
A.D. King, D. Vassiliev,
Torsion waves in metric-affine field theory,
Class. Quantum Grav. 18 (2001) 2317--2329.

\bibitem{exact solutions}
D. Kramer, H. Stephani, E. Herlt, M. MacCallum,
Exact Solutions of Einstein's Field Equations,
Cambridge University Press, Cambridge, 1980.

\bibitem{mielkepseudoparticle}
E.W. Mielke,
On pseudoparticle solutions in Yang's theory of gravity,
Gen. Rel. Grav. 13 (1981) 175--187.

\bibitem{Nakahara}
M. Nakahara,
Geometry, Topology and Physics,
Institute of Physics Publishing, Bristol, 1998.

\bibitem{peres}
A. Peres,
Some gravitational waves,
Phys. Rev. Lett. 3 (1959) 571--572.

\bibitem{peresweb}
A. Peres,
abstract to preprint hep-th/0205040
(reprinting of \cite{peres}).

\bibitem{griffiths3}
P. Singh, J.B. Griffiths,
A new class of exact solutions of the vacuum quadratic
Poincar\'e gauge field theory,
Gen. Rel. Grav. 22 (1990) 947--956.

\bibitem{kiev}
D. Vassiliev,
A metric-affine field model for the neutrino,
in:  Noncommutative Structures in Mathematics and Physics
(eds. S. Duplij and J. Wess),
Kluwer Academic Publishers, Dordrecht, 2001, pp. 427--439.

\bibitem{pseudo}
D. Vassiliev,
Pseudoinstantons in metric-affine field theory,
Gen. Rel. Grav. 34 (2002) 1239--1265.

\bibitem{weylquadraticaction}
H. Weyl,
Eine neue Erweiterung der Relativit\"atstheorie,
Ann. Phys. 59 (1919) 101--133.


\bibitem{yang}
C.N. Yang,
Integral formalism for gauge fields,
Phys. Rev. Lett. 33  (1974) 445--447.

\end{thebibliography}
\end{document}